\documentclass[sn-mathphys-num]{sn-jnl}

\usepackage{graphicx}%
\usepackage{multirow}%
\usepackage{amsmath,amssymb,amsfonts}%
\usepackage{amsthm}%
\usepackage{mathrsfs}%
\usepackage[title]{appendix}%
\usepackage{xcolor}%
\usepackage{textcomp}%
\usepackage{manyfoot}%
\usepackage{booktabs}%
\usepackage{algorithm}%
\usepackage{algorithmicx}%
\usepackage{algpseudocode}%
\usepackage{listings}%
\usepackage{pdflscape}
\usepackage{adjustbox}
\usepackage{caption}
\usepackage{rotating}
\usepackage{colortbl}
\usepackage{float}
\usepackage{placeins}
\usepackage{verbatim}
\usepackage{geometry}

\theoremstyle{thmstyleone}%
\theoremstyle{thmstyletwo}%

\theoremstyle{thmstylethree}%

\begin{document}

\title[Article Title]{Head Kinematics and Brain Tissue Deformation from Soccer Heading: A Review of Implications for Brain Injury Risk}

\author[1]{\fnm{Christopher} \sur{Lewis}}\email{cjlst102@mail.rmu.edu}

\author[1]{\fnm{Anu} \sur{Tripathi}}\email{tripatha@rmu.edu}

\author[2]{\fnm{Alison} \sur{Brooks}}\email{brooks@ortho.wisc.edu}

\author[3]{\fnm{Peter} \sur{Ferrazzano}}\email{ferrazzano@pediatrics.wisc.edu}

\author[4]{\fnm{Joseph} \sur{Andrews}}\email{joseph.andrews@wisc.edu}

\author[5]{\fnm{Traci} \sur{Snedden}}\email{traci.snedden@cuanschutz.edu}

\author[4]{\fnm{Christian} \sur{Franck}}\email{cfranck@wisc.edu}

\author*[1]{\fnm{Rika} \sur{Carlsen}}\email{carlsen@rmu.edu}

\affil*[1]{\orgdiv{Department of Engineering}, \orgname{Robert Morris University}, \orgaddress{\city{Moon Township}, \state{PA}, \country{USA}}}

\affil[2]{\orgdiv{Department of Orthopedics and Rehabilitation}, \orgname{University of Wisconsin--Madison}, \orgaddress{\city{Madison}, \state{WI}, \country{USA}}}

\affil[3]{\orgdiv{Waisman Center}, \orgname{University of Wisconsin--Madison}, \orgaddress{\city{Madison}, \state{WI}, \country{USA}}}

\affil[4]{\orgdiv{Department of Mechanical Engineering}, \orgname{University of Wisconsin--Madison}, \orgaddress{\city{Madison}, \state{WI}, \country{USA}}}

\affil[5]{\orgdiv{College of Nursing}, \orgname{University of Colorado Anschutz Medical Campus}, \orgaddress{\city{Aurora}, \state{CO}, \country{USA}}}




\abstract{\textbf{Purpose:} 
Repeated heading of soccer balls has raised concerns of potential long-term neurological effects. 
Consequently, numerous studies have estimated 
head kinematics and 
brain deformation due to soccer headers across different cohorts and play scenarios  to identify higher risk conditions. 
However, heterogeneity in study design, data collection, and analysis has produced inconsistent findings, and injury risk is infrequently reported. Therefore, a meta-analysis of the existing literature was conducted to identify knowledge gaps and inform future studies assessing injury risk in soccer. 

\textbf{Methods:}  We synthesized data from studies reporting head kinematics or brain deformation from soccer headers on human subjects. The data from these studies were analyzed to obtain the risk of mild traumatic brain injury (mTBI) based on applicable injury metrics and risk curves. 

\textbf{Results:} The meta-analysis revealed specific trends, indicating that match scenarios, corner and goal-kicks, top and oblique impacts, and older age cohorts were associated with higher head kinematics, while sex-based trends were inconclusive. The choice of sensor system affected the estimated head kinematics, with headband sensors consistently measuring higher kinematics than mouthpiece sensors.
The data showed large variability stemming from heterogeneous study designs, 
limiting the applicability of the observed trends. 
These factors also influenced injury risk predictions, with estimated concussion risks generally below 20\%. 

\textbf{Conclusion:} This review reveals trends in mTBI risk from soccer heading across different cohorts and play scenarios. 
It also underscores the need for standardized reporting of kinematics and brain deformation to enable mTBI risk estimation and meaningful cross-study comparisons.

}



\maketitle

\section{Introduction}\label{sec1}

Soccer, or football, when compared to other contact sports, is unique in that intentional head impacts, most commonly via heading the ball, are a fundamental part of play and occur frequently once players reach the age where heading is permitted under current regulations \cite{langdon2022heading}. Although concern for the long-term neurological health of contact sport athletes continues to grow, the question of whether soccer heading poses a significant risk of brain injury remains unanswered. Unlike sports such as American football, which have been studied much more extensively to inform the development of safety equipment, such as helmets \cite{newman2005verification}, soccer is typically played without helmets or other protective measures for the head, as the large majority of impacts appear to be sub-concussive \cite{kroshus2015concussion, o2014evaluation}. However, the frequency of these impacts, especially among youth players \cite{beaudouin2020uefa}, has impelled researchers to question whether the cumulative effect of these sub-concussive events could contribute to long-term neurological impairment \cite{mohamed2025impact, grijalva2023hyper,urbanik2024changes}. 

Despite its public perception as a lower-risk sport, soccer is among the leading causes of sports-related traumatic brain injuries (TBIs), particularly in youth and female athletes \cite{gessel2007concussions, covassin2012role}. Players often underestimate the risk due to the absence or infrequency 
of overt, concussive collisions seen in sports like football or hockey \cite{chrisman2013qualitative}. However, emerging epidemiological data suggest that repetitive sub-concussive impacts from heading may contribute to subtle but meaningful neurological changes over time \cite{koerte2012white}. It is worth noting, however, that some studies have found no significant association between heading exposure and cognitive impairment \cite{rodrigues2019no}, underscoring the lack of consensus on the long-term effects of sub-concussive impacts in soccer.

Addressing the link between head impacts in soccer and neurological outcomes requires the improved understanding of the biomechanics behind heading a soccer ball. The most common way that researchers have sought to characterize these biomechanics is through the measurement of head kinematics using wearable sensors and the estimation of brain deformation using computational simulations 
for different types of headers across various demographics of players \cite{basinas2022systematic}. Although several studies have measured linear head accelerations and/or angular head accelerations and velocities during heading, the reported peak magnitudes vary markedly, often due to methodology inconsistencies \cite{snowden2021heading}. 
Furthermore, very few computational studies have been performed to obtain the extent of brain deformation resulting from soccer headers. 

\begin{figure}[h!]
    \centering
    \includegraphics[width=1.1\linewidth]{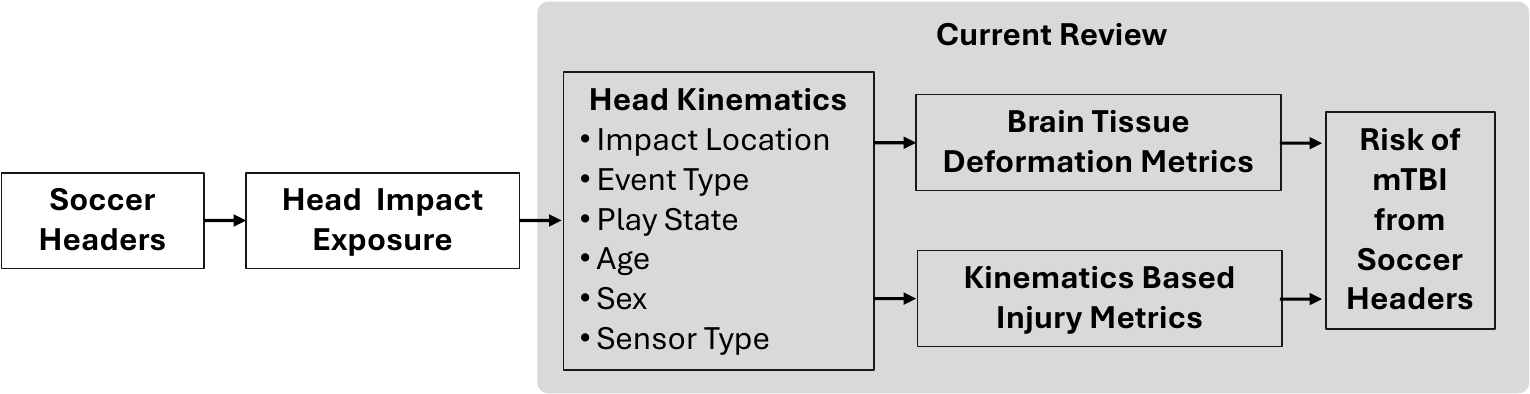}
    \caption{Flowchart summarizing the scope of the literature review.}
    \label{fig:Overview}
\end{figure}

This meta-analysis synthesizes experimental studies with controlled, scripted headers; observational studies capturing realistic, live play; and simulation-based studies
on soccer heading by human subjects, and evaluates whether the data supports a meaningful risk of mild traumatic brain injury (mTBI). 
We compute  mTBI risk using synthesized data and injury risk metrics proposed in the literature.
The review is organized thematically.  The methods section (Section \ref{kin_def}) defines key parameters used in the meta-analysis.  Section \ref{kinematics} focuses on head kinematics studies, and Section \ref{strain} focuses on computational studies.  An assessment of the risk of mTBI from soccer heading is provided in Section \ref{res:mtbiRisk}, followed by a discussion of limitations in Section \ref{lim} and a summary of future directions in Section \ref{future}.  The overall scope of this literature review is shown in Figure \ref{fig:Overview}.


\section{Methods}
\label{kin_def}

Multiple experimental studies have collected soccer heading kinematic data across various demographics of players, and several computational studies have assessed the brain deformation resulting from these headers.  We identified studies through targeted searches of the PubMed and Google Scholar databases, and citation chaining from relevant papers. 
Only studies reporting head kinematics or brain deformation metrics associated with human soccer heading were included. 
In cases where the reported data were insufficiently detailed or did not align with our analysis requirements, study authors were contacted to obtain additional data breakdowns. 
Studies that reported only the number of head impacts without providing the head kinematics were not included in the meta-analyses. Given the inconsistency in header type definitions, data reporting, wearable sensors, and computational models used across the studies, we define the key variables in this section 
to facilitate meaningful comparison and highlight how differences in reporting complicate direct comparisons. 


\FloatBarrier
\subsection{Study Methodologies}

Studies examining head kinematics in soccer heading span a wide range of methodologies, from highly controlled laboratory setups to field-based observations and biomechanical simulations. This variability in experimental design, while valuable for exploring different aspects of heading biomechanics, introduces limitations when attempting to compare results across studies or synthesize injury risk. 
All of the included studies gathered kinematic data by one of the three methodologies as defined in Table \ref{tab:methodology}. 

\begin{table}[h!]
\centering
\caption{Study methodologies included in this review
}
\begin{tabular}{p{0.25\linewidth} p{0.7\linewidth}}
\toprule
\textbf{Methodology} & \textbf{Description} \\
\midrule
Laboratory-Based Studies & 
Controlled heading drills using standardized ball delivery systems to isolate variables such as ball velocity or header type. Head kinematics were recorded using wearable sensors. \\[6pt]

Field-Based Studies & 
Players heading balls in real training or match environments, using wearable sensors to collect head kinematic data during actual play. \\[6pt]

Simulated Impact Studies & 
Head kinematics obtained directly from finite element head models that simulate the soccer ball impact to the head.\\

\bottomrule
\end{tabular}
\label{tab:methodology}
\end{table}

\FloatBarrier
\subsection{Header Types}
\paragraph{Impact Locations}
Studies have compared head kinematics resulting from different soccer ball impact locations on the head during a header \cite{huber2023neurophysiological,kenny2022head,harriss2019head,bretzin2017sex,miller2020characterizing,delang2025acute}, usually as a result of different techniques. 
We categorized each header into one or more of the comparisons described in Table \ref{tab:technique}.  The headers recorded during a throw-in were assumed to be frontal unless the study suggested otherwise.

\begin{table}[h!]
\centering
\caption{Soccer ball impact locations on the head during a soccer header.}
\begin{tabular}{p{0.2\linewidth} p{0.8\linewidth}}
\toprule
\textbf{Technique} & \textbf{Description} \\
\midrule
Frontal & 
Contact made with the forehead; typically associated with direct, oncoming impacts.\\

Oblique & 
Contact near the temple or side of the head, often from lateral or off-axis ball trajectories, and usually accompanied by an intentional rotation of the head to aid in redirecting the ball.\\

Top & 
Impact to the crown or top of the head, usually from vertical challenges or poorly timed jumps.\\
\bottomrule
\end{tabular}
\label{tab:technique}
\end{table}


\FloatBarrier
\paragraph{Play States}

Headers were grouped by the type of play that initiated the ball trajectory.  These header types are summarized in Table \ref{tab:playstate}. 

\begin{table}[h!]
\centering
\caption{Header types based on the type of play that initiated the ball trajectory.}
\begin{tabular}{p{0.2\linewidth} p{0.8\linewidth}}
\toprule
\textbf{Play State} & \textbf{Description} \\
\midrule
Goal Kick & 
Headers resulting from long-distance kicks taken from the goal area, typically high, direct trajectories.\\

Corner Kick & 
Headers initiated from a cross delivered from the corner arc, often involving aerial challenges in front of goal.\\

Throw-in & 
Headers following a ball reintroduced by an overhand throw from the sideline, usually shorter trajectories with less speed.\\

Free Kick & 
Headers responding to set plays from a stationary kick outside the corner or goal area, often involving organized positioning.\\

Live Ball & 
Headers during open play not initiated by a set piece; includes dynamic, unstructured trajectories.\\

Drill & 
Headers occurring in structured, repetitive, or non-gameplay-based training scenarios.\\
\bottomrule
\end{tabular}
\label{tab:playstate}
\end{table}
\FloatBarrier
\paragraph{Event Types}

Lastly, as described in Table \ref{tab:eventtype}, headers were categorized based on the event type or the setting in which they occurred. 

\begin{table}[h!]
\centering
\caption{Categorization of headers by event type 
}
\begin{tabular}{p{0.2\linewidth} p{0.8\linewidth}}
\toprule
\textbf{Event Type} & \textbf{Description} \\
\midrule
Match & 
Headers performed during official games or scrimmages.\\

Training & 
Headers performed during practices, structured drills, or lab-based experiments.\\
\bottomrule
\end{tabular}
\label{tab:eventtype}
\end{table}

This classification system allowed for the extraction and comparison of kinematic variables across studies despite differences in terminology and reporting standards. The following sections provide a consolidated overview of the kinematic variables most commonly measured in these studies and highlight how differences in reporting complicate direct comparisons.

\subsection{Kinematic Measures }\label{subsec3}

Across the studies reviewed, head impact severity during soccer heading is most commonly quantified using three primary kinematic variables: peak linear acceleration (PLA), peak angular velocity (PAV), and peak angular acceleration (PAA). These metrics represent the mechanical response of the head to ball impact and form the basis for assessing potential injury metrics. 

Linear acceleration, typically measured in gravitational units (g), reflects the translational movement of the head immediately following impact. Most soccer heading studies report relatively low PLA values, often below 50 g, which all fall below thresholds commonly associated with acute concussion \cite{mcintosh2014biomechanics}. However, the implications of repeated sub-concussive exposure at these levels remain unclear. 

Angular velocity (rad/s) and angular acceleration (rad/s²) describe the rotational dynamics of the head. These measures are important because rotational motions are considered more relevant than linear ones in the context of concussive injury and diffuse axonal  injury \cite{meaney2011biomechanics,gennarelli1982biomechanics}. Angular velocity is more commonly reported than angular acceleration, 
since most wearable sensors rely on gyroscopes that directly measure angular velocity, requiring angular acceleration to be derived through additional computation. 

Not all studies report kinematic values in a standardized format. Some provide component values (x, y, z) along individual anatomical axes (e.g., coronal, sagittal, axial), while others only report resultant magnitudes. When available, resultant measures were extracted for comparison across studies. If only axis-specific values were given, vector magnitudes were reconstructed when possible. Studies lacking sufficient information for reconstruction were excluded from those specific comparisons. 

\subsection{Wearable Sensors}\label{sensors}

Accurate measurement of head kinematics during soccer heading depends heavily on the type, placement, and coupling of the sensors used. Across the studies reviewed, a range of wearable devices were deployed, including headbands, mouthguards, and adhesive skin patches, each with distinct trade-offs in accuracy, usability, and ecological validity.

Instrumented mouthguards and mouthpieces offer superior skull coupling and have demonstrated higher accuracy in capturing true head kinematics 
\cite{stitt2021laboratory,liu2020validation}. However, 
they face some practical barriers in implementation, especially at youth and amateur levels, likely due to the time, cost, and logistical effort required to custom-fit each mouthguard to an individual athlete. 

Headband-mounted sensors were among the most commonly used, particularly in field-based studies. These sensors offer ease of deployment but have been shown to suffer from skull-sensor decoupling and overestimation of angular velocity and acceleration due to headband 
displacement relative to the skull \cite{huber2021laboratory}. Despite these limitations, they remain popular due to their plug-and-play usability, requiring minimal subject-specific fitting or calibration, which allows for rapid deployment across large cohorts. 

Adhesive patches placed behind the ears have also been used in some lab-based settings. While these offer better coupling than loose headbands, they are still vulnerable to relative motion between the device and skull due to soft tissue motion, particularly during high-speed or off-axis impacts \cite{wu2016vivo}.

Sensor sampling rate, filtering algorithms, and trigger thresholds vary widely across studies and are often under-reported, making it difficult to directly compare reported values. For example, some devices apply proprietary filtering that smooths or clips high-magnitude events, while others may include low-level noise as valid impacts. This variability further complicates cross-study interpretation and contributes to the inconsistency of reported kinematic outcomes. Despite these limitations, wearable sensors remain the only practical solution for large-scale data collection in soccer heading research. However, future studies must prioritize transparent reporting of sensor specifications, validation protocols, and data processing methods to improve reproducibility and interpretability across the field.

To improve reproducibility and comparability across the field, 
recently published Consensus Head Acceleration Measurement Practices (CHAMP) guidelines outline the best practices for the validation, reporting, and analysis of head acceleration measurements \cite{gabler2022consensus}. These recommendations emphasize independent laboratory validation of wearable devices using ATDs, standardized reference sensor systems, and transparent reporting of data processing protocols. Alignment with these practices will be critical for advancing the reliability of wearable sensor data in soccer heading research. 

In our analyses, we label each data source to identify the sensor type -- mouthpiece, headband, or skin patch -- to account for the bias introduced by the choice of sensor mount.

\subsection{Meta Analysis}\label{stat}

We calculated the means and standard deviations of the peak kinematics reported in the reviewed papers to study the effects of the five most studied factors: play state, impact location, event type, age, and sex. 
Since headband-mounted sensors have been shown to produce higher kinematic measurements and variability compared to mouthpiece-based sensors, which have superior skull coupling and reduced motion artifacts, we assess the effect of each factor separately for each sensor type. 
To evaluate the statistical significance of a trend, we used Welch's t-test, where $p<0.05$ is considered significant and $p<0.005$ is highly significant. 
It should be noted that in cases where substantial variability in study designs leads to multi-modal distributions, the reliability of this test is reduced.

\section{Head Kinematics}
\label{kinematics}



In this section, we summarize the trends from the head kinematics studies, including the demographics (by age and sex) covered across the studies (Section \ref{kin:dem}) and the 
kinematic outcomes according to key comparative dimensions: impact location (Section \ref{res:location}), event type (Section \ref{res:event}), play state (Section \ref{res:state}), player age (Section \ref{res:age}), and sex (Section \ref{res:sex}). 

\FloatBarrier
\subsection{Demographic Coverage}\label{kin:dem}

\begin{figure}[h!]
    \centering
    \includegraphics[width=1\linewidth]{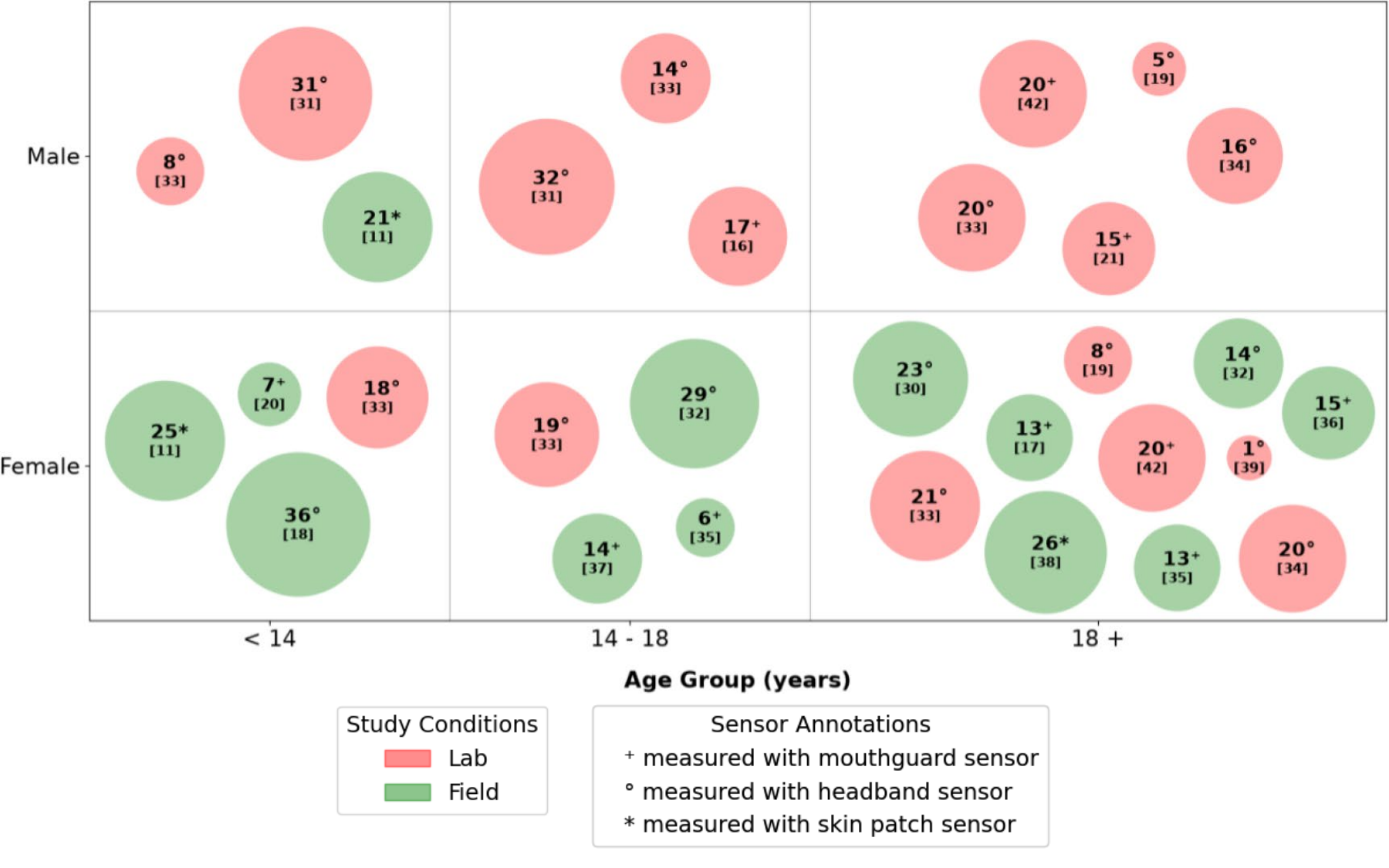}
    \space
    \caption{Distribution of soccer head impact kinematic-only studies across age and sex demographics. Each bubble represents a distinct study cohort; the number of participants analyzed per cohort is displayed inside each bubble and corresponds to the bubble size. Bubble color indicates whether the study was conducted in a laboratory or field setting, as determined by a review of each study’s methodology.  The superscript symbol represents the type of wearable sensor used in the study, and the reference number of the study is listed in brackets.   
    }
    \label{Demographic bubbleplot}
\end{figure}

Figure \ref{Demographic bubbleplot} highlights several demographic and methodological patterns within the 18 included studies reporting head kinematics from soccer heading. The majority of studies focus on participants aged 18 and older, which were collegiate players (16 cohorts), compared to 14 cohorts involving athletes under 18. 
While studies are almost evenly split between field-based (10 studies) and lab-based settings (8 studies), lab-based data covers a wider range of cohorts, highlighting notable gaps in field studies on older ($>$ 14 years) male cohorts. 
In terms of measurement tools, instrumented mouthguards and headbands are more frequently used (8 studies each) than skin patches (2 studies). The studies with a larger number of participants typically used a headband or a skin patch \cite{chrisman2013qualitative, harriss2019head, lamond2018linear, kalichova2019soccer, mccuen2015collegiate, caccese2018sex, nowak2020neuro}.

\subsection{Effect of impact location}\label{res:location}

\begin{figure}[htbp]
    \centering
    \includegraphics[width=0.95\linewidth]{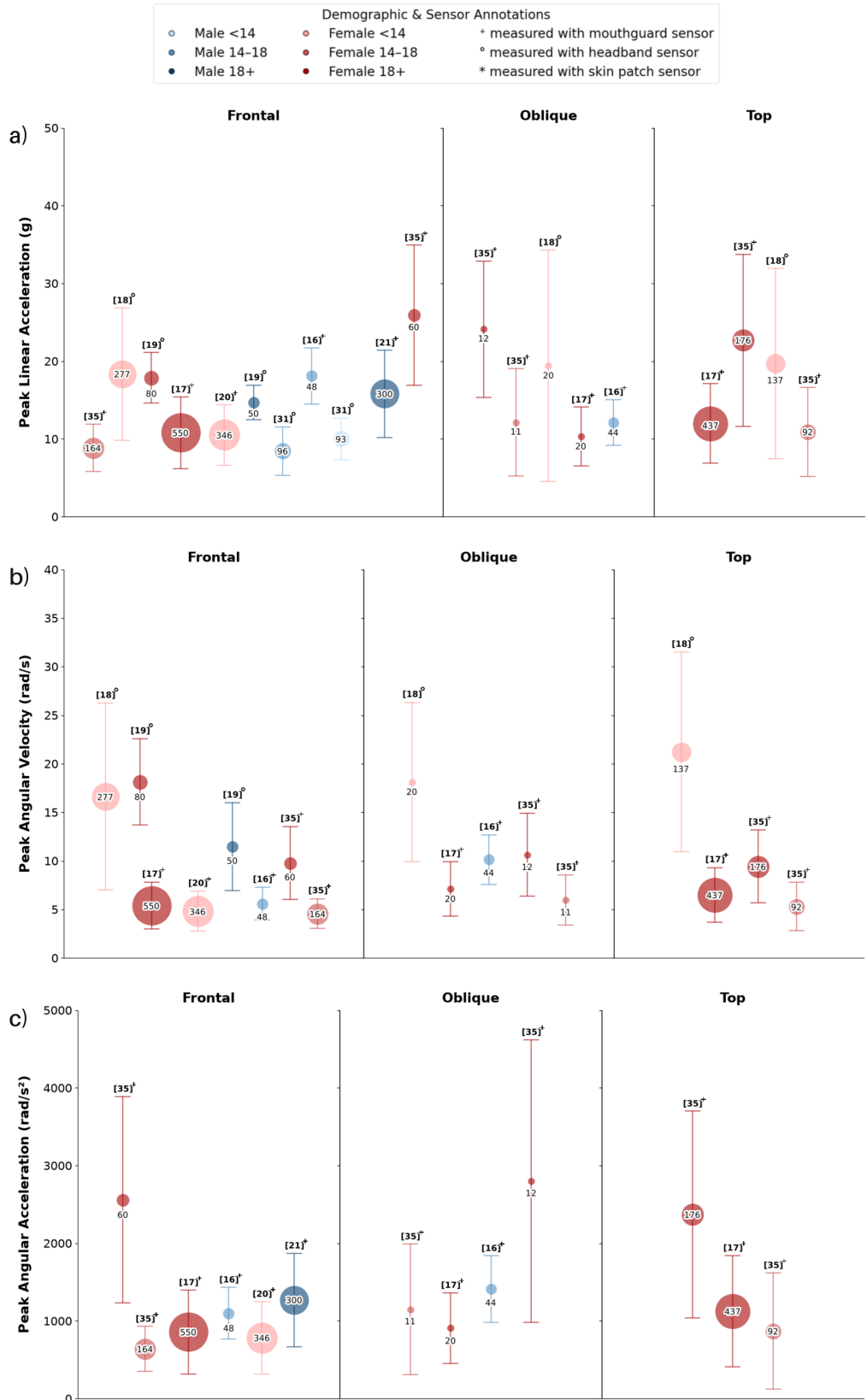}
    \space
    \caption{Peak head kinematics recorded across different soccer header impact locations, as reported in multiple studies: (a) peak linear acceleration, (b) peak angular velocity, and (c) peak angular acceleration. 
    Marker size is scaled to the number of headers observed in each cohort, with the number of headers displayed near each marker. Error bars indicate the standard deviation.}
    \label{Combined Technique}
\end{figure}

Multiple studies have investigated how the impact location of a soccer ball on the head affects the head kinematics. Figure \ref{Combined Technique} shows the range of peak linear acceleration (PLA), peak angular velocity (PAV), and peak angular acceleration (PAA) values that have been recorded from soccer heading studies  for frontal, oblique (side), or top impact locations. 

Individual mouthguard studies investigating the effect of impact locations found conflicting trends, with Huber et al. finding frontal impacts to cause significantly higher translational and lower rotational kinematics compared to oblique impacts \cite{huber2023neurophysiological}, while data from Kenny et al. did not show any significant difference between frontal, oblique, and top impacts \cite{kenny2022head, smith2026influence}. 
Combining the data from mouthguard studies, a significantly larger number of frontal impacts was recorded ($n = 1468$), compared to oblique ($n = 87$) or top ($n = 705$) impacts. The combined PLA data shows statistically similar PLA for all impact locations (frontal impacts: mean $\pm$ std $= 15.0 \pm 13.1$ g; top impacts: $15.2 \pm 13.5$ g; and oblique impacts: $14.7 \pm 12.2$ g) (Figure \ref{Combined Technique}a). 
On the other hand, the lowest PAV was found in frontal impacts ($6.0\pm5.4$ rad/s), followed by top ($7.1\pm5.3$ rad/s) and oblique ($8.5\pm6.2$ rad/s) impacts (all statistically significant, $p<0.05$), suggesting that impacts not aligned with a frontal trajectory generate greater rotational motion (Figure \ref{Combined Technique}b). 
Only mouthguard studies reported PAA (Figure \ref{Combined Technique}c). 
Frontal impacts produced relatively lower mean PAA ($1202\pm 1684$ rad/s$^2$) compared to top impacts ($1455\pm1687$ rad/s$^2$) ($p<0.05$). The oblique impacts ($1567\pm 2100$ rad/s$^2$) showed high variability, resulting in low statistical difference from frontal or top impacts ($p>0.1$). 

Given the small number of headband studies reporting data on frontal \cite{harriss2019head, bretzin2017sex}, oblique, and top impacts \cite{harriss2019head}, our meta-analysis found the same trends as the individual studies \cite{harriss2019head}.
The frontal impacts resulted in the lowest PLA ($n = 596$, PLA $= 13.9 \pm 10.23$ g) when compared to the oblique ($n=20$, PLA $=19.4 \pm 14.9$ g) and top ($n=137$, PLA $= 19.69 \pm 12.2$ g) impacts. While top impacts resulted in the highest PAV ($21.21 \pm 10.27$ rad/s), followed by the oblique ($18.11 \pm 8.2$ rad/s) and frontal impacts ($15.4 \pm 11.5$ rad/s), all trends are statistically insignificant ($p>0.1$) (Figure \ref{Combined Technique}a). 


\subsection{Effect of Event Type}\label{res:event}
\begin{figure}[htbp]
    \centering
    \includegraphics[width=0.95\linewidth]{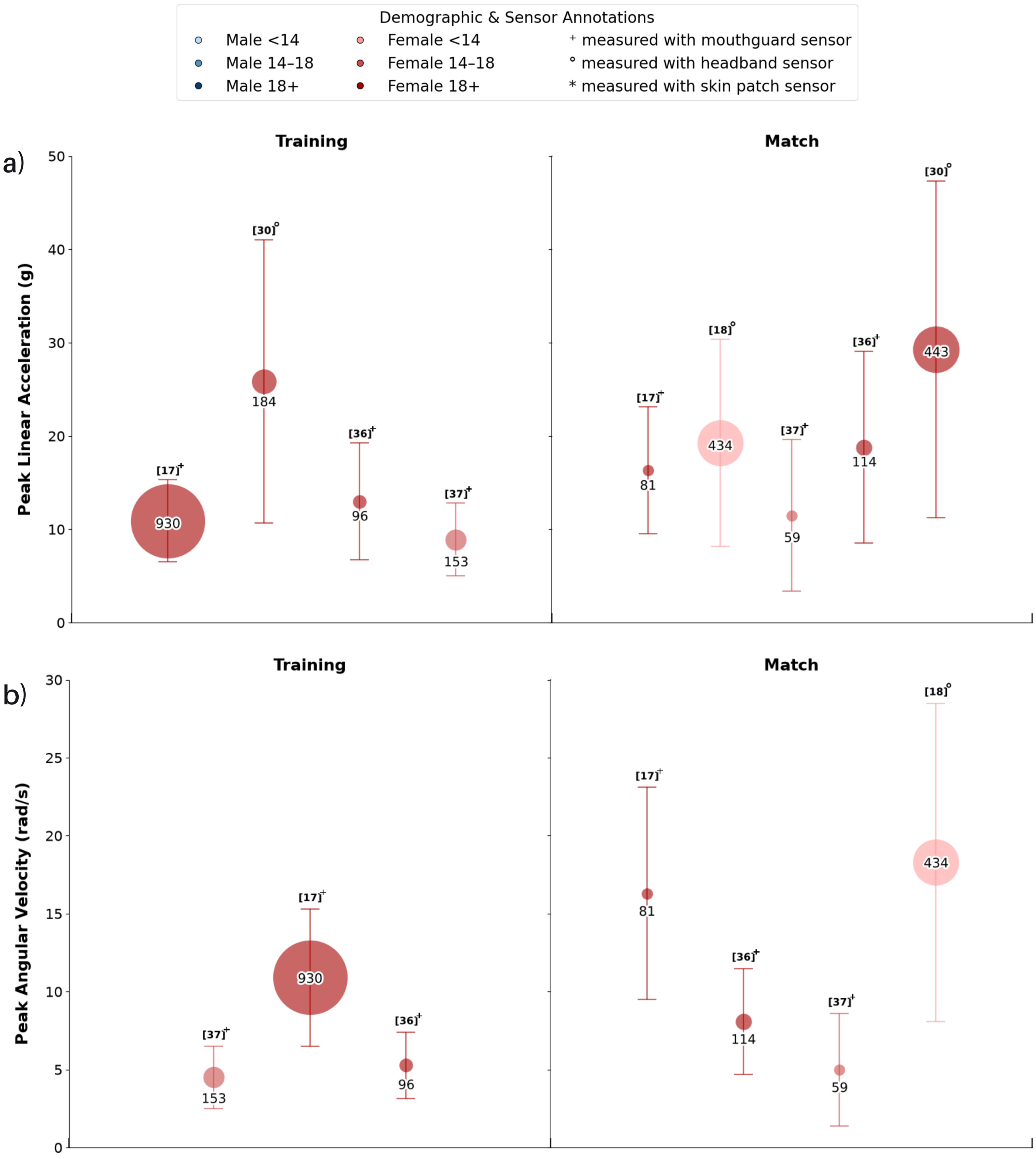}
    \space
    \caption{Peak head kinematics recorded during training and match headers across multiple studies: (a) peak linear acceleration and (b) peak angular velocity. Marker size is scaled to the number of headers observed in each cohort, with the number of headers displayed near each marker. Error bars indicate the standard deviation.}
    \label{Combined Event Type}
\end{figure}

Across all included studies, head kinematics varied systematically by event type, with match play consistently producing higher measured values than training across all sensor types (Figure \ref{Combined Event Type}). The observation from the meta-analysis is consistent with the individual studies \cite{filben2021comparison,lamond2018linear,kenny2022head,filben2024assessing, press2017quantifying}. 
This pattern was most apparent in PLA, where match-play cohorts exceeded their training counterparts in every study (Figure \ref{Combined Event Type}a) ($p<0.005$). PAV measurements, though fewer in number, followed the same trend (Figure \ref{Combined Event Type}b): match-play impacts yielded higher rotational speeds than those recorded during training ($p<0.005$). It should be noted that the dataset is dominated by female participants aged 18 and older, with relatively few male or youth cohorts represented, limiting broader generalization across sex and age groups. 

\subsection{Effect of Play State}\label{res:state}

\begin{figure}[htbp]
    \centering
    \includegraphics[width=0.95\linewidth]{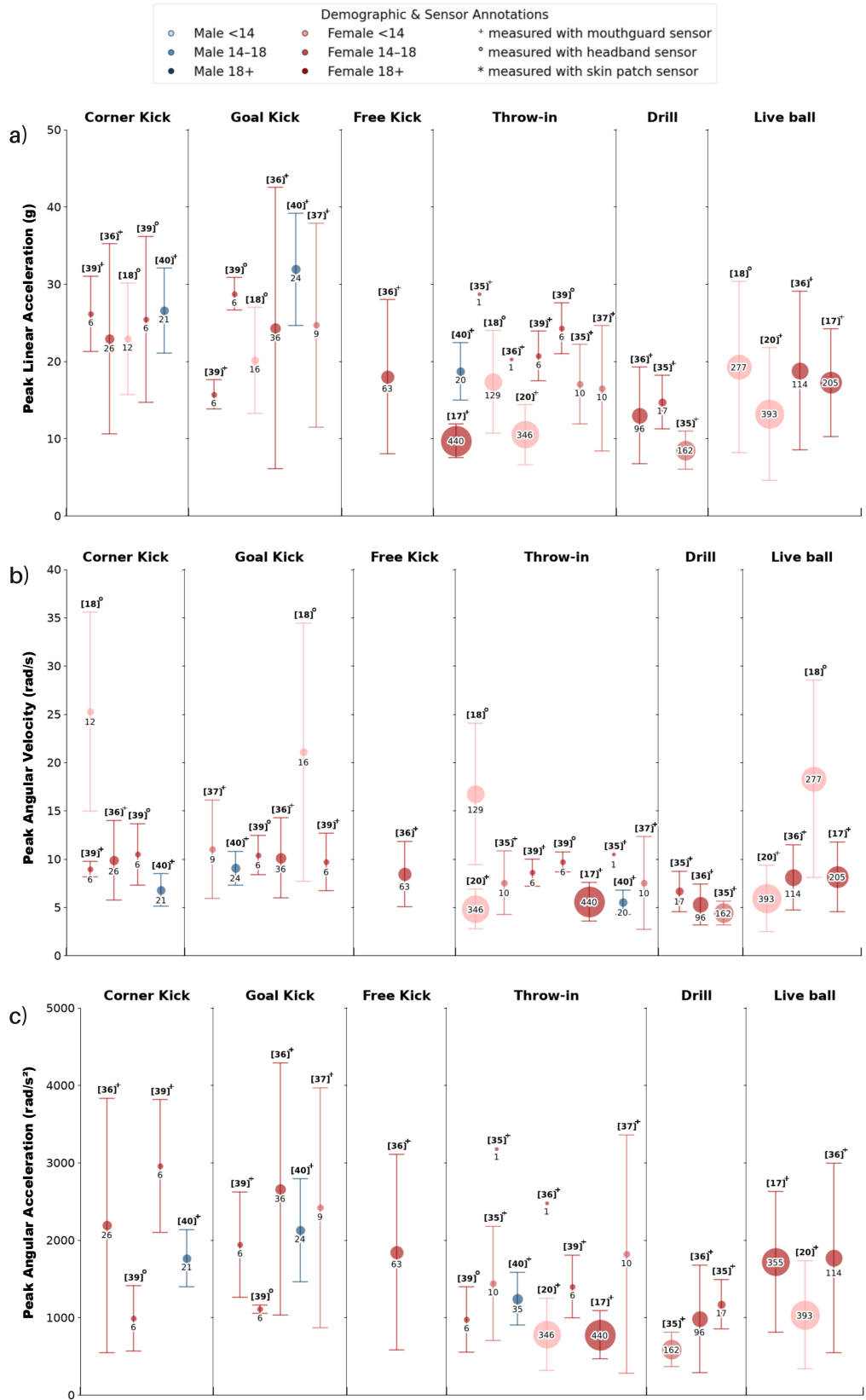}
    \space
    \caption{Peak head kinematics recorded during different soccer play states (corner kicks, goal kicks, throw-ins, drills, and live play), as reported in multiple studies: (a) peak linear acceleration, (b) peak angular velocity, and (c) peak angular acceleration. Marker size is scaled to the number of headers observed in each cohort, with the number of headers also displayed on or just below each marker. Error bars indicate the standard deviation.}
    \label{Combined Play State}
\end{figure}

 Studies have explored how different play states that initiate the ball trajectory leading to a header influence the resulting head kinematics (Table \ref{tab:playstate}).    
Among the mouthguard data, PLA vary markedly by play state, with corner kicks ($n=53$, $25.2 \pm 14.3$ g) and goal kicks ($n=75$, $24.2 \pm 23.7$ g) producing the highest PLA; free-kicks ($n=63$, $18.0 \pm 10.0$ g), throw-ins ($n=834$, $17.7 \pm 11.7$ g), and live ball scrimmages ($n=712$, $16.4 \pm 15.1$ g) producing PLAs in the moderate range; and practice drills produced the smallest PLA ($n=275$, $12.1 \pm 7.6$ g). 
Unlike the other comparisons where the headband-measured values were higher than the mouthguard-measured values, this trend was not consistent here, 
underscoring that play state factors can outweigh sensor type in certain contexts. These patterns are illustrated in Figure \ref{Combined Play State}a.
The headband data also yielded larger PLA in corner ($n=18$, $24.2 \pm 12.9$ g) and goal kicks ($n=22$, $24.4 \pm 7.2$ g) than live ball ($n=227$, $19 \pm 11$ g) and throw-ins ($n=135$, $20.8 \pm 7.4$ g) \cite{harriss2019head,tripathi2025field}.  There were no headband data on free kicks and practice drills.

Among the mouthguard studies, the highest PAVs were in the corner ($n=53$, $8.5 \pm 4.5$ rad/s) and goal ($n=75$, $10.0 \pm 7.4$ rad/s) kicks, while throw-ins ($n=833$, $7.1 \pm 6.7$ rad/s) and drill-based headers ($n=275$, $5.4 \pm 3.2$ rad/s) produced the lowest values. However, no significant difference was observed between corner and goal kicks. The headband data, on the other hand, showed larger PAV in live-ball scrimmages ($n=227$, $18.3 \pm 10.2$ rad/s) than corner kicks ($n=18$, $17.9 \pm 10.8$ rad/s), goal kicks ($n=22$, $15.7 \pm 13.5$), and throw-ins ($n=135$, $13.2 \pm 7.4$). Note that the large spread in headband data led to bimodal distributions (due to significantly different data processing and performance of the headband), and the observed trends were not significant (all $p>0.1$). 

The mouthguard PAA data indicate that throw-ins ($n=849$, $1640 \pm 1875$ rad/s$^2$), drill-based headers ($n= 275$, $914 \pm 797$ rad/s$^2$), and live-play ($n=862$, $1507\pm 1676$ rad/s$^2$) generally produced lower values when compared to corner ($n=53$, $2303 \pm 1891$ rad/s$^2$) and goal kicks ($n=75$, $2286 \pm 2442$ rad/s$^2$) (Figure \ref{Combined Play State}c). 

Overall, both corner and goal kicks produced the largest peak head  kinematics, while drill-based headers were the lowest ($p<0.05$). Throw-ins, live-ball, and free-kick headers fell in the intermediate range. However, while categorizing headers into discrete “play states” can help describe the context of impact, these classifications are inherently limited. 
The actual kinematic profile due to a header is determined by the specific ball–player interaction at the moment of impact, not the nominal play state in which it occurs.
Although physical factors that govern head kinematics, such as specific ball speeds, trajectories, or techniques, are statistically more likely to be involved in a specific play state, these are only broad generalizations, and consequently, play state-based classification risks oversimplifying and potentially misrepresenting the mechanical demands and injury risk of heading in soccer. 


\subsection{Effect of Age}\label{res:age}

Age is considered a potentially important determinant of head impact kinematics. Individual studies found conflicting trends: Kalichová and Lukášek \cite{kalichova2019soccer} reported a linear decrease in PLA with increasing age, suggesting that younger players may be more vulnerable to high-magnitude head impacts, while McCuen et al. \cite{mccuen2015collegiate} found high school players sustained significantly lower average PLA compared to adult players, and PAA did not differ significantly between groups. 

An overall meta-analysis on mouthguard data shows increasing PLA with age group: the younger age group ($<14$ yr) experienced the lowest PLA ($n=1075$, $11.4 \pm 10.2$ g), followed by high school players ($14-18$ yrs) ($n= 770$, $14.4 \pm 23.5$ g), with the highest PLA found in the oldest age group ($>18$ yr) ($n=4056$, $19.1 \pm 40.3$ g) (all $p<0.005$). Note that even though the large spread resulted in multi-modal distributions, as evidenced by negative values of the mean minus the standard deviation, these trends were found to be highly significant.  
The same trend was observed for PAV and PAA: players in the younger age group ($<14$ yr) experienced the lowest PAV ($5.2 \pm 4.5$ rad/s) and PAA ($867 \pm 960$ rad/s$^2$), followed by high school players ($14-18$ yrs) ($6.8 \pm 10.7$ rad/s and $1367 \pm 2960$ rad/s$^2$), and then the older age group ($>18$ yr) ($8.5 \pm 16.3$ rad/s and $1725 \pm 4687$ rad/s$^2$) (all $p<0.005$). 
The trend is most apparent in Figures {\ref{Combined Technique}a-b and \ref{Combined Play State}b across mouthguard studies.  Younger players, particularly female athletes under age 14, have frequently sustained lower peak angular velocities and peak linear accelerations than their older counterparts \cite{miller2020characterizing,kenny2022head}. 

A similar trend was found in the headband data with the younger ($<14$ yr) age group experiencing lower PLA ($n=1138$, $18.5 \pm 29$ g) and PAV ($n=1045$, $19.4 \pm 28.5$ rad/s) as compared to the older ($>18$ yr) age group ($n=775$, $23.7 \pm 26.5$ g and $n=148$, $12 \pm 7.4$ rad/s) ($p<0.005$). Figure \ref{Combined Event Type}a also shows lower PLA in headband studies of female athletes under age 14 during match events compared to their older counterparts \cite{harriss2019head,lamond2018linear}. 



\subsection{Effect of Sex}\label{res:sex}

Across studies, sex-related differences in head kinematics are not consistent across all kinematic measures. 
The meta-analysis on all mouthguard data shows higher PLA in males ($n=457$, $20.5 \pm 12.3$ g) than in females ($n=5444$, $16.7 \pm 46.1$ g), while headband data shows lower PLA in males ($n=239$, $11.1 \pm 4.7$ g) than in females ($n=1770$, $22.0 \pm 39.2$ g) (all $p<0.005$). Collectively, the mouthguard data showed comparable mean PAVs between males ($n= 157$, $7.4 \pm 4.1$ rad/s; $n=472$, $1486 \pm 1163$ rad/s$^2$) and females ($n=5443$, $7.8 \pm 19.6$ rad/s; $n=6183$, $1582 \pm 5504$ rad/s$^2$), although the female data showed greater variability. Since the data is collected across a wide range of play states and impact locations, we stratify the comparison between male and female head kinematics based on these factors to remove the confounding effects arising from differences in header distributions.

The mouthguard data for both frontal and oblique impacts (Figure \ref{Combined Technique}a-b), shows higher PLA, PAV, and PAA in females \cite{kenny2022head, miller2020characterizing} than in males \cite{huber2023neurophysiological,delang2025acute} (Figure \ref{Combined Technique}c). 
Similarly, Figure \ref{Combined Play State}a-c shows higher PLA, PAV, and PAA in female mouthguard data for corner-kicks, goal-kicks, and throw-ins \cite{filben2021comparison, tripathi2025field}, as compared to the data for males \cite{barnes2024investigation}. However, fewer studies on male cohorts can bias these comparisons due to other confounding factors, such as age and event types.


\paragraph{Anatomical and Anthropometric Factors}
Several studies have investigated how differences in anatomy and anthropometry between sexes may contribute to sex-related differences in head kinematics.  In a controlled lab study using headbands, Caccese et al. \cite{caccese2018sex} found that females exhibited significantly higher peak linear and angular accelerations than males when performing frontal headers. The paper speculates that this may be due to the smaller head mass and lower neck strength of the average female when compared to the average male. Müller and Zentgraf \cite{muller2021neck} similarly observed greater head impact magnitudes in female players during laboratory-based heading drills using headbands, attributing the difference in part to lower neck strength. Bretzin et al. \cite{bretzin2017sex} reported that adult females display lower neck girth and neck strength compared to men, and experience increased head impact kinematics during soccer heading. Most recently, Abbasi Ghiri et al. \cite{abbasi2025exploring} reported that individuals with smaller head mass, more typical of female athletes, experienced significantly higher PLA and PAA during a laboratory-based heading drill, but head mass had no significant effect on PAV. 


\paragraph{Contextual and Methodological Limitations}

Importantly, all studies directly comparing male and female head kinematics were conducted in laboratory settings rather than during live matches or training scenarios \cite{caccese2018sex, muller2021neck, bretzin2017sex, abbasi2025exploring}. In these controlled environments, ball velocities are standardized across participants, regardless of sex, which does not necessarily reflect real-world play. Sakamoto \cite{sakamoto2014comparison} reported significantly greater ball velocities from male players during instep kicks. This discrepancy implies that, in real-world play, female athletes are likely to head slower-moving balls, meaning that the elevated accelerations seen in lab trials may not translate to match or training conditions for women. 
So, while findings are consistent across laboratory protocols, the absence of field-based comparisons leaves a critical gap in understanding sex-related differences under realistic match conditions. Future work should prioritize matched male–female cohorts, standardized measurement tools, and field-based data collection that captures the range of ball velocities and play dynamics typical of each sex. 

\section{Brain Tissue Deformation}
\label{strain}

Previous studies have correlated brain tissue deformation metrics with the risk of mTBI. Brain deformation is typically estimated from computational simulations of head impact events conducted through finite element (FE) analysis.  While stress-based deformation metrics have been proposed in early studies \cite{zhang2004proposed, oeur2015comparison}, more recent studies have focused on strain-based measures \cite{wright2012axonal, gabler2016assessment, wu2021evaluation}. Some of these strain-based metrics include the peak maximum principal strain (MPS) in the brain tissue, which is the maximum tensile deformation relative to the tissue’s undeformed state \cite{gabler2016assessment, wu2021evaluation, zhang2024objective}, and the 
peak maximum axonal strain (MAS), which is the peak tensile strain component along the fiber tract direction in the white matter of the brain \cite{wright2012axonal}. 
A volumetric damage, or cumulative strain damage measure (CSDM), has also been proposed, which is defined as the volume fraction of brain tissue that exceeds an injury strain threshold \cite{takhounts2013development}. Since there is little consensus on the injury strain threshold for mTBI, strain values ranging from 0.10 to 0.25 have been used for CSDM \cite{perkins2022assessment, takhounts2013development}. Corresponding strain rate measures have also been proposed, including maximum principal strain rate (MPSR), maximum axonal strain rate (MASR), cumulative strain rate damage measure (CSRDM), as well as a combination of strain and strain rate measures \cite{wu2021evaluation, miller2021brain}.

Among finite element (FE) studies that have characterized  brain tissue deformation resulting from soccer headers, the peak von Mises stress, MPS, and CSDM have been reported.  
MPS has been reported as either the mean value across different regions of interest \cite{brooks2021purposeful} or the 95th percentile value of the whole brain (MPS95) \cite{barnes2024investigation, filben2021header, huber2024finite}. 
To simulate a head impact and estimate the deformation metrics, researchers often use measured head kinematics data as inputs into FE models of the human head, which are typically developed from image-based anatomical geometries. 
The quality of the output is therefore contingent on the biofidelity of the FE model, the accuracy of the kinematic input, and the computational methods used for simulation. 


The five studies included in this review that estimated tissue deformation from soccer heading employed a wide range of methodologies 
as summarized in Table \ref{tab:computational_studies}. Across these studies, there was substantial variability in the sensor type used to obtain the head kinematic data, including instrumented mouthguards (Prevent Biometrics IMM, custom-fitted iMG systems, custom mouthpiece-mounted accelerometers and gyroscopes) \cite{barnes2024investigation, filben2021header, huber2024finite}, headbands (GForce Tracker) \cite{brooks2021purposeful}, and in one case, no physical sensor was used.  Instead, the kinematic inputs were generated via simulated impacts \cite{perkins2022assessment}. 
Kinematic inputs were obtained from field-based play, laboratory setups, and purely computational simulations, with cohorts differing in age (youth: $<14$ yrs, adolescent: 14–18 yrs, adult: $>$18 yrs) and sex (male and female). The number of headers analyzed per study ranged from 65 in controlled lab experiments to 483 in large field datasets. 

The FE simulations were performed on different head models, including the Kungliga Tekniska Högskolan (KTH) model \cite{kleiven2007predictors}, scaled Global Human Body Models Consortium (GHBMC) \cite{mao2013development}, and the Atlas-Based Brain Model (ABM) \cite{miller2016development}, which run on the solver LS-DYNA (Ansys Inc.), as well as the 
University College Dublin Brain Trauma Model (UCDBTM) \cite{horgan2003creation} and a custom 50th-percentile male head model \cite{perkins2022assessment}, which runs on Abaqus software (Simulia, Dassault Systems). The FE head models differed in 
the level of anatomical detail, 
FE mesh size (mesh refinement), and material models, each of which has been shown to affect computed injury metrics \cite{ho2009can, tripathi2025effect, miller2017validation, zhao2018material, giudice2019analytical}. The header studies also varied in whether the models were scaled to the cohort head size, which has also been reported to affect brain deformation \cite{giudice2023investigating}. Finally, the studies differed in terms of the injury metrics reported: two studies reported von Mises stress \cite{perkins2022assessment,huber2024finite}, four studies reported 95th percentile peak MPS \cite{perkins2022assessment, huber2021laboratory, filben2021header, barnes2024investigation}, one study reported average peak MPS in different ROIs \cite{brooks2021purposeful}, and two studies reported CSDM with a strain injury threshold of 0.10 \cite{barnes2024investigation} and 0.05 \cite{perkins2022assessment}. 
We discuss the results from these computational studies on soccer heading in light of these differences in the following section.


\subsection{Demographic Coverage}\label{comp:dem}

The computational studies that simulate soccer headers using measured head kinematics data from specific sex and age cohorts are summarized in Figure \ref{Computational demographic bubbleplot}. We find that computational studies are relatively scarce within the current literature, especially compared to the number of experimental studies collecting soccer heading kinematics data. 
Most subjects featured in the computational studies were high school-aged females, with all female cohorts studied in field settings and all male cohorts studied in laboratory environments. Critically, there were no computational studies involving players under the age of 14, highlighting a major gap in understanding how heading affects the youngest athletes. 

\begin{figure}[htbp]
    \centering
    \includegraphics[width=0.9\linewidth]{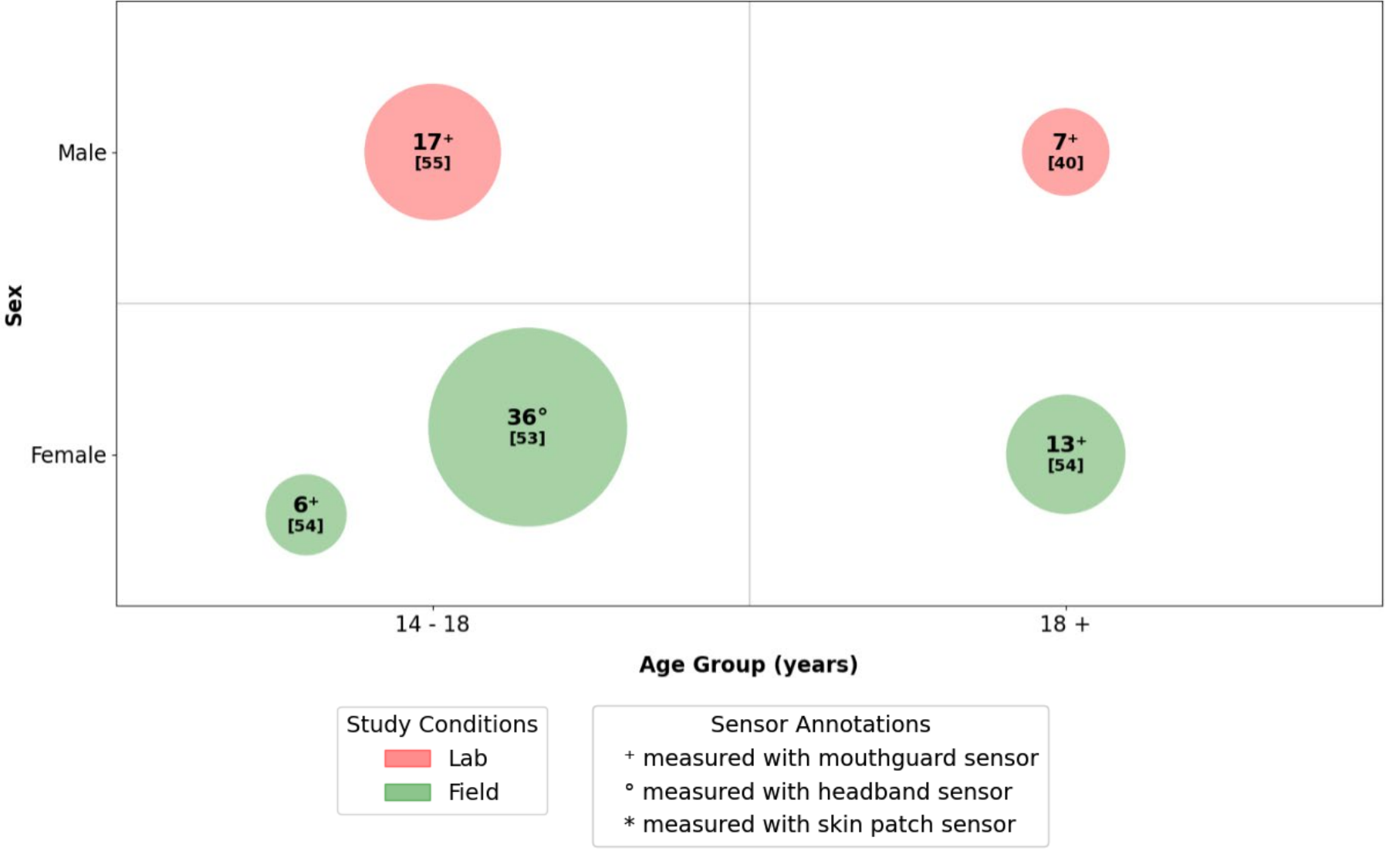}
    \space
    \caption{Distribution of soccer header computational 
    studies across age and sex demographics. Each marker represents a distinct study cohort; the number of participants analyzed per cohort is displayed inside each marker and corresponds to the marker size. Marker color indicates whether the study was conducted in a laboratory or field setting, as determined by a review of each study’s methodology.} 
    \label{Computational demographic bubbleplot}
\end{figure}


\subsection{Observed Trends}\label{comp:trends}




\paragraph{Effect of Head Kinematics on Brain Strain}
Despite the substantial methodological variation across studies, several patterns emerge. Within each study, higher MPS values tended to coincide with greater peak angular velocities and accelerations, reinforcing the established relationship between rotational motion and shear strain in brain tissue \cite{carlsen2021quantitative}. However, this trend does not persist in cross-study comparisons.
In Perkins et al., the head kinematic inputs were generated via simulated impacts, and this study predicted the highest PAA ($5017-11030$ rad/s$^2$) but lowest MPS estimates ($0.0016-0.044$) (100th percentile MPS) \cite{perkins2022assessment}, as compared to the simulations based on sensor kinematics data  that reported higher MPS ($0.048-0.10$) (95th percentile MPS) for lower PAA values (mean PAA: $655-2230$ rad/s$^2$) \cite{filben2021header, barnes2024investigation, huber2021laboratory}. Among studies based on sensor data, there are also inconsistencies with the expected trend. For example, Filben et al. \cite{filben2021header} reported an MPS95 of 0.048 for a PAA of 655 $\pm$ 435 rad/s$^2$, while Huber et al. \cite{huber2021laboratory} predicted a lower MPS95 of 0.039 for a higher PAA of 1147 $\pm$ 45 rad/s$^2$.  These results highlight the effect that computational methods (e.g., choice of head model) and the kinematics time history can have on the strain results. 


\paragraph{Effect of Age and Sex}
Only one study directly compared age groups under similar field conditions, reporting higher MPS in adult females (0.073) than in youth females (0.048) \cite{filben2021header}. However, the study did not account for the effect of head size in their model, using the same head size (adult male) for both populations. Larger head sizes, typically in older cohorts, have been found to result in slightly higher strains for the same kinematic input \cite{giudice2023investigating}.  Scaling the head models to the appropriate head size would have increased the difference between the younger and the older cohorts. When comparing the study on the young male cohort \cite{huber2024finite} with the study on the slightly older male cohort \cite{barnes2024investigation}, we find that the frontal impacts generated lower PAV (5.5$\pm$0.2 rad/s), PAA (1147$\pm$45 rad/s$^2$), and MPS95 (0.039) in the younger cohort as compared to the older cohort   (PAV: 7.2$\pm$2.2, PAA: 1730$\pm$2.18, MPS95: 0.096). Both studies used an adult male FE head model; head size scaling in the younger cohort study would have further reduced the MPS95 values.  
No study directly compared male and female players within the same experimental setup, instead sex-specific findings came from separate methodologies, sensor types, computational models, and reported deformation metrics, making it difficult to isolate true sex effects.

\paragraph{Effect of Impact Location}
Huber et al. \cite{huber2024finite} analyzed the effect of impact location using mouthguard head kinematic data and the KTH head model.  They 
predicted a lower MPS95 of 0.039 for frontal impacts with a PAV of $5.5\pm0.2$ rad/s and a slight increase in MPS95 (0.042) for oblique impacts with almost double the PAV ($10.2\pm0.4$). 
Similarly, the simulated impact study by Perkins et al. \cite{perkins2022assessment} found the lowest peak MPS for frontal impacts (0.0018), with the highest peak MPS occurring for side impacts (0.044).

Overall, while MPS offers a more direct tissue-level indicator of mechanical brain loading than raw kinematic data, its interpretation is currently constrained by heterogeneity in modeling approaches, sensor technologies, and the validity of input kinematics.


\newgeometry{left=1cm, right=1cm, top=1cm, bottom=2cm, landscape}

\begin{sidewaystable}[htbp]
\caption{Summary of methodologies and findings from five computational studies that directly estimated maximum principal strain (MPS) during soccer heading. Reported variables include sensor type, finite element (FE) brain model, cohort demographics, peak linear and angular kinematics, and corresponding MPS values (mean, 95th, or 100th percentile).}
\label{tab:computational_studies}

\setlength{\tabcolsep}{6pt}
\renewcommand{\arraystretch}{3}
\definecolor{lightgray}{gray}{0.85}
\definecolor{medgray}{gray}{0.85}

\rowcolors{2}{gray!10}{white}
\arrayrulecolor{lightgray}
\newcolumntype{P}[1]{>{\raggedright\arraybackslash}p{#1}}
\begin{tabular}{|P{0.02\linewidth}|P{0.07\linewidth} |P{0.08\linewidth} |P{0.08\linewidth} |P{0.03\linewidth} |P{0.09\linewidth} |P{0.09\linewidth} |P{0.09\linewidth} |P{0.06\linewidth} |P{0.11\linewidth}}
\arrayrulecolor{lightgray}  
\toprule
\rowcolor{medgray}
\textbf{Ref.} & \textbf{Sensor Type} & \textbf{FE Model Used} & \textbf{Cohort Age (years)} & \textbf{Sex} & \textbf{PLA (g)} & \textbf{PAV (rad/s)} & \textbf{PAA (rad/s\textsuperscript{2})} & \textbf{Reported Metric(s)} & \textbf{Value(s)} \\
\midrule

\cite{huber2024finite} & Prevent Biometrics Mouthguard & KTH \newline (Adult male) & 13 -- 18 & Male &
Frontal: \newline 17.4 $\pm$ 0.5 \newline Oblique: \newline12.1 $\pm$ 0.4 &
Frontal: \newline5.5 $\pm$ 0.2 \newline Oblique: \newline10.2 $\pm$ 0.4 &
Frontal: \newline1147 $\pm$ 45 \newline Oblique: \newline1410 $\pm$ 65 &
95th percentile MPS &
Frontal: 0.039 \newline Oblique: 0.042 \\[6pt]

\cite{brooks2021purposeful} & GForce Tracker Headband & GHBMC \newline (Scaled to 13F)& 13.4 $\pm$ 0.9 & Female &
16.1 [12.3--22.1] &
16.2 [10.5--21.2] &
Not reported &
Region-specific mean MPS &
Corpus callosum: 0.102 \newline Thalamus: 0.083 \newline Brainstem: 0.048 \\[6pt]

\cite{filben2021header} & Wake Forest Mouthpiece & ABM \newline (Adult male) &
Youth: \newline 15.27 $\pm$ 0.11 \newline Adult: \newline 20.19 $\pm$ 1.34 & Female &
Youth: \newline9.46 $\pm$ 4.54 \newline Adult: \newline22.5 $\pm$ 14.9 &
Youth: \newline 4.64 $\pm$ 2.07 \newline Adult: \newline9.03 $\pm$ 5.37 &
Youth: \newline655 $\pm$ 435 \newline Adult: \newline2230 $\pm$ 1890 &
95th percentile MPS &
Youth: \newline 0.0477 \newline Adult:\newline 0.0728 \\[6pt]

\cite{barnes2024investigation} & iMG Mouthguard & UCDBTM \newline (Adult male) & 20.1 $\pm$ 1.0 & Male &
26 $\pm$ 7.9 &
7.20 $\pm$ 2.18 &
1730 $\pm$ 611 &
95th percentile MPS &
0.0962 \\[6pt]

\cite{perkins2022assessment} & FE Impact Simulation (no physical sensor) & Custom adult male head model & N/A & N/A &
Front: 86.0 \newline Side: 108.2 \newline Top: 69.5 \newline Rear: 135.4  &
Not reported &
Front: 5643.9 \newline Side: 11030.0 \newline Top: 5017.4 \newline Rear: 10140.0 &
100th percentile MPS &
Front: 0.0018 \newline Side: 0.044 \newline Top: 0.0041 \newline Rear: 0.038 \\
\bottomrule
\arrayrulecolor{black}  
\end{tabular}
\end{sidewaystable}

\restoregeometry

\section{Implications for TBI Risk}\label{res:mtbiRisk}

The risk of concussion from head impact events can be estimated using either head kinematics-based or brain tissue deformation-based measures (Sections \ref{kinematics} and \ref{strain}) \cite{macmanus2022material, rowson2022review, gabler2016assessment}. Although repeated sub-concussive head impacts in soccer have raised concerns of potential accumulation over time causing long-term neurodegeneration, it is currently not possible to measure this accumulation of damage. Therefore, in this review, we focus on the potential risk of concussion from a single head impact event using  reported head kinematics (Section \ref{injuryKinematics}) and brain deformation data (Section \ref{injuryBrainDef}).   


\subsection{Kinematics-Based Injury Metrics}\label{injuryKinematics}

Multiple head kinematics-based injury metrics have been defined and 
correlated to concussion risk \cite{rowson2012rotational}. These injury metrics are often derived by performing logistic regression analysis on concussion data from sports or animal tests \cite{takhounts2013development, rowson2012rotational, kleiven2006evaluation}.  No kinematic injury metric is universally accepted for all types of head impacts, with some injury metrics performing better for specific scenarios, such as motor vehicle collisions, pedestrian impacts, or American football impacts \cite{gabler2016assessment, zhan2021relationship}. 
Since an injury metric specifically for soccer headers has yet to be be proposed, we currently need to rely on existing kinematics-based injury metrics to assess concussion risk from soccer heading. 

Most kinematics-based injury metrics are defined as a function of the peak head kinematic measures, such as PAV, PAA, or PLA \cite{gabler2016assessment}.  Some also account for the duration of the head impact event, such as the severity index (SI), head injury criterion (HIC), brain injury threshold surface (BITS), and head impact power (HIP) \cite{gabler2016assessment,versace1971review,gadd1966use}.  Furthermore, some metrics are based on resultant kinematic measures whereas others account for the head rotation direction, requiring the kinematics along the primary anatomical axes (axial, sagittal, and coronal) to be known.  Some of these directionally dependent metrics include the brain injury criterion (BrIC) \cite{takhounts2013development, takhounts2011kinematic} and rotational velocity change index (RVCI) \cite{yanaoka2015investigation}. 
Since most soccer heading kinematic studies report only the resultant peak kinematic values for each header, only a limited number of injury metrics that are based soley on resultant peak kinematics can be evaluated. Furthermore, while previous studies have proposed injury metrics based on translational acceleration (e.g., HIC), more recent work has shown that angular kinematics is the primary driver of brain tissue deformation \cite{zhang2006role}. Given these factors, we have chosen to estimate the risk of concussion using two injury metrics based on resultant angular kinematics: 
(1) the Brain Injury Criterion (BrIC$_{\mathrm{R}}$), which was initially developed to assess injury from occupant crash tests \cite{takhounts2013development}; and (2) an angular velocity concussion risk function derived from American football data \cite{rowson2012rotational}.  The risk of concussion is assessed with these metrics using the head kinematics data presented in Section \ref{kinematics}.

Multiple versions of the Brain Injury Criterion (BrIC) have been proposed over the years.  For this analysis, we use a simplified scalar form of BrIC based on the resultant angular velocity and defined as follows:
\begin{equation}
\text{BrIC}_{\mathrm{R}} = \frac{\omega_{\max}}{\omega_{c}},
\end{equation}
\noindent where $\omega_{\max} = \text{max}\left(\sqrt{\omega_x^2 + \omega_y^2 + \omega_z^2}\right)$ is the resultant angular velocity, and $\omega_{c} = 54.85~\text{rad/s}$ is the critical resultant angular velocity \cite{takhounts2013development}.  A concussion risk (AIS 2) curve equation is provided in \cite{takhounts2013development}, which is based on a logistic regression  between BrIC$_{\mathrm{R}}$, MPS, and concussion outcomes. 
It should be noted that the most widely used version of BrIC accounts for the directional dependence of angular velocity \cite{gabler2016assessment, perkins2022assessment, abbasi2025exploring, huber2023neurophysiological}, and this version has been shown to have significantly higher correlation (R$^2\sim$0.84) with injury risk compared to the version of BrIC used here based on resultant angular velocity alone, i.e., BrIC$_{\mathrm{R}}$ (R$^2\sim0.64$) \cite{takhounts2013development}. Other variations of the directional BrIC have also been proposed \cite{laituri2016new, gabler2018development}. Given that only the resultant head kinematics are reported for the majority of existing soccer header studies, the directional BrIC cannot be applied to the data derived from these studies.

In addition to the Brain Injury Criterion, we also estimate concussion risk based on the angular velocity injury risk values provided by Rowson et al. \cite{rowson2012rotational}. This concussion injury risk function was later revised to incorporate the effects of both translational and angular accelerations \cite{rowson2013brain}. However, the updated metric could not be applied to the soccer heading data since the translational and angular acceleration are not consistently reported for each head impact, limiting our analysis to the angular velocity alone. 

The results for the two injury risk metrics based on the PAV soccer heading data  from Section \ref{kinematics} are presented in Figure \ref{Combined BrIC}. The primary y-axis (left) shows the mTBI risk (AIS 2) computed from Takhounts et al. \cite{takhounts2013development}, denoted as \emph{Takhounts' Injury Risk Probability}. The background shading and the accompanying color scale on the secondary y-axis (right) correspond to the concussion risk from Rowson et al. \cite{rowson2012rotational}, denoted as \emph{Rowson's Concussion Risk Probability}.

\subsubsection{Injury Risk Analysis based on Head Kinematics}

\begin{figure}[htbp]
    \centering
    \includegraphics[width=.95\linewidth]{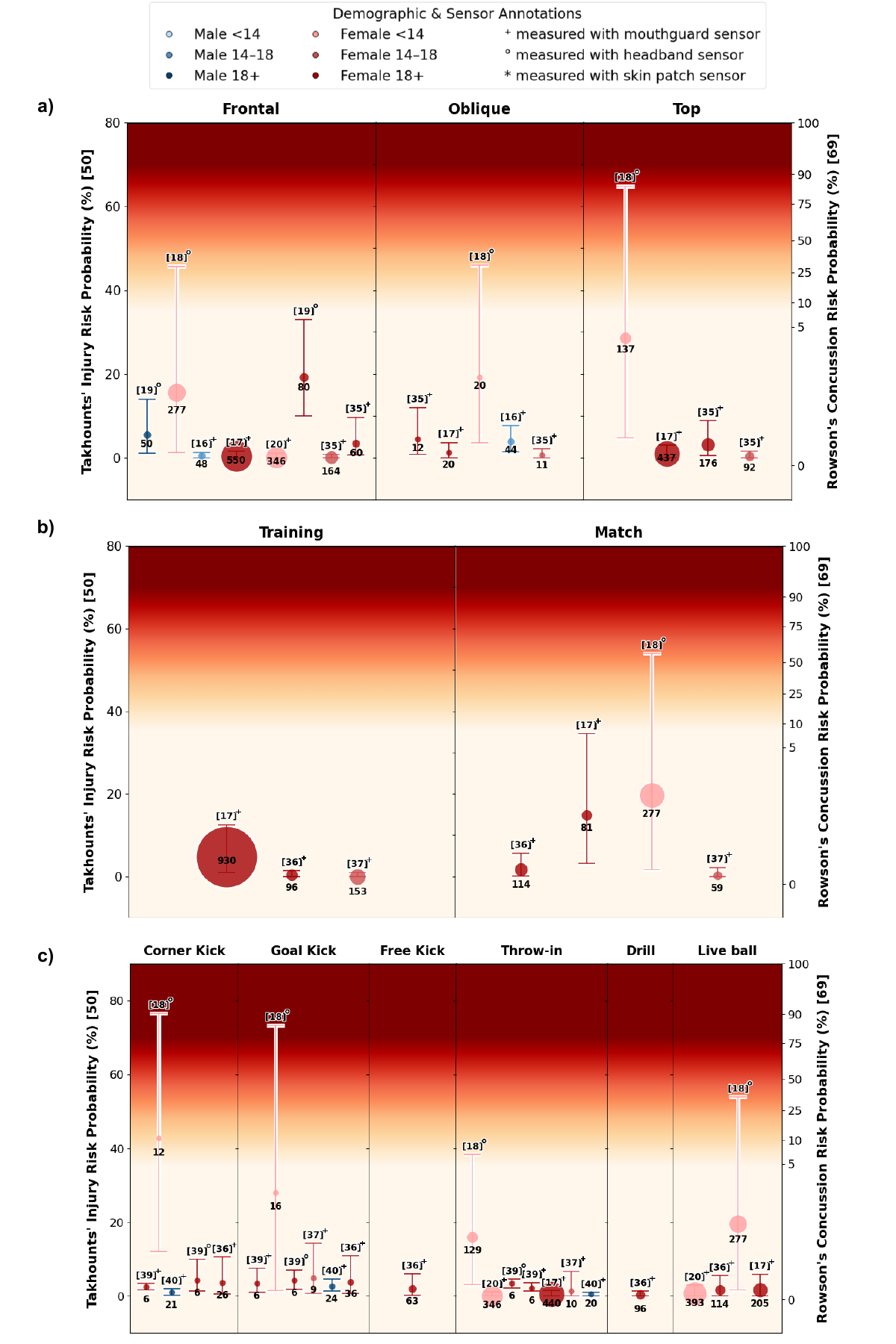}
    \space
    \caption{Estimated brain injury risk associated with soccer heading across multiple contexts: (a) different heading techniques, (b) training versus match play, and (c) different play states. The primary y-axis (left) shows the injury risks calculated based on the correlation provided between BrIC\textsubscript{R}, MPS, and concussion by Takhounts et al. 2013 \cite{takhounts2013development}. The background shading and the accompanying color scale on the secondary y-axis (right) correspond to the injury risk based on the correlation provided between PAV and concussion by Rowson et al. 2012 \cite{rowson2012rotational}. Marker size is scaled to the number of headers observed in each cohort, with the number of headers also displayed on or just below each marker. Error bars indicate the standard deviation of the measured values. }
    \label{Combined BrIC}
\end{figure}

\paragraph{Effect of Impact Location}
Since both injury risk metrics used in this study are based on monotonous functions of peak angular velocity, they follow the same trends as the head PAV. 
We find that differences across impact locations were small based on mouthguard data, where frontal, oblique, and top headers result in $<10\%$ risk of injury based on both risk metrics \cite{takhounts2013development, rowson2012rotational} (Figure \ref{Combined BrIC}a). The headband data \cite{harriss2019head,bretzin2017sex} results in a mean injury risk of about 20\%  for frontal and oblique impacts, and about 30\% for top impacts based on Takhounts et al. (BrIC\textsubscript{R}) \cite{takhounts2013development}, while being $<10\%$ based on Rowson et al. \cite{rowson2012rotational}. The upper bound of the standard deviation of the headband data reaches up to 70 - 90\% based on both risk metrics \cite{takhounts2013development, rowson2012rotational} (Figure \ref{Combined BrIC}a). It should be noted that the known issue of skull-headband decoupling reduces the confidence in the headband-based concussion risk estimates.  
When comparing these results against other soccer heading studies that have reported an estimated concussion risk based on impact location, the results are varied.  A simulated impact study found the BrIC-based injury risk to be below 5\% for top impacts, 41\% for rear impacts, 48\% for side impacts, and 23\% for frontal impacts \cite{perkins2022assessment}.  A mouthguard study, on the other hand, found the BrIC-based injury risk to be $<5$\% for frontal impacts \cite{barnes2024investigation}. The risk of injury based on HIC \cite{hayes2007forensic} was reported to be below 5\% for front, side, rear, and top impacts in both simulated impact and mouthguard studies \cite{perkins2022assessment, barnes2024investigation}.

\paragraph{Effect of Event Type}
Differences in Takhounts' injury risk (BrIC\textsubscript{R}) values between event types (training and match) were modest but consistently higher during match play (Figure \ref{Combined BrIC}b). The mouthguard-based training impact data showed $<15$\% injury risk based on Takhounts' metric and $<5$\% based on Rowson's metric \cite{kenny2022head, filben2021comparison,filben2024assessing}. On the other hand, mouthguard-based match header data resulted in up to 30\% risk based on Takhounts' metric and $<10\%$ risk based on Rowson's metric. 
Several match-play cohorts in the instrumented headband studies produced standard deviations extending to around 50\% injury risk based on both Takhounts' and Rowson's Injury Risk Probability \cite{harriss2019head}.
Comparing the studies that are similar in sensor mounts (mouthguards), cohort age, and sex (adult female) \cite{kenny2022head, filben2021comparison, filben2024assessing}, it can be concluded that match headers generally have a higher risk of injury. 

\paragraph{Effect of Play State}
The risk of concussion for most play states was below  $10$\% based on both sensor mounts (mouthguard and headband) and risk metrics \cite{takhounts2013development, rowson2012rotational} (Figure \ref{Combined BrIC}c). However, one headband study on a female  cohort ($<14$ yrs old) exhibited large variability \cite{harriss2019head}, with the upper bound of the standard deviation reaching an injury risk of $>90$\% based on the Rowson metric and $75$\% based on Takhounts' metric. 


Overall, typical soccer heading exposures generally correspond to a low injury risk probability range ($<10\%$ concussion risk), but the choice of sensor mounts has a significant effect on calculated injury risks. Certain high-velocity headers, such as corner and goal kicks, particularly during matches, introduce variability that can push injury risk values into higher risk bands. Also, while the concussion risk values lie below $10\%$, repeated instances can accumulate to a higher probability, which cannot be quantified based on available data and injury metrics. 
Therefore, low injury risk values should not be interpreted as evidence that heading is benign; rather, they highlight the need for injury metrics calibrated to the unique magnitude, frequency, and biomechanics of soccer-specific head impacts.

While the above analysis provides useful insight into injury risk in soccer, it carries important limitations. Many available injury risk metrics could not be utlized due to the lack of reported data on header duration and head kinematics components along principal directions, which have been shown to be important when estimating injury risk \cite{takhounts2013development}. Therefore, future studies should report header durations as well as kinematics components and should provide this data for each measured impact. 
Injury metrics derived from sensor-reported data \cite{rowson2012rotational} or computational models \cite{takhounts2013development} also remain subject to modeling and measurement errors. 


\subsection{Tissue Deformation-Based Injury Metrics}\label{injuryBrainDef}

In addition to kinematics-based injury metrics, the risk of concussion can also be estimated from brain deformation measures, such as MPS and CSDM. 
As described in Section \ref{strain} and shown in Figure \ref{InjuryRisk}a, the brain deformation from soccer headers can be estimated from FE simulations using detailed head models \cite{huber2024finite,brooks2021purposeful, perkins2022assessment, barnes2024investigation, filben2021header}. Both tissue strain and strain rate have been shown to play an important role in cellular injury, and deformation-based injury risk curves have been derived from both these measures \cite{gonzalez2024cortical,estrada2021neural,bar2016strain}. Since FE simulations are computationally expensive that often take hours or even days to run, several methods have also been proposed to obtain real-time estimates of brain deformation and injury risk, including reduced-order models derived from FE parametric studies, such as the Rate Inclusive Strain Estimate (RISE) \cite{carlsen2021quantitative}; second-order mechanical analog models such as the Universal Brain Injury Criterion (UBrIC) \cite{gabler2018development} and Diffuse Axonal Multi-Axis General Evaluation (DAMAGE) \cite{gabler2019development}; analytical wave-dynamics solutions of idealized head models \cite{wan2023finite}; and machine-learning based predictions \cite{zhan2021rapidly, zhan2024adaptive, zhan2022finding, arani2026novel}. 

Two approaches have been utilized to link brain deformation to brain injury risk. The first approach is based on logistic regression between the computed brain deformation from simulated head impact events and reported concussion outcomes (injury vs no-injury) \cite{gabler2016assessment, zhang2024objective, takhounts2013development}. While logistic regression studies provide injury risk curves, their accuracy can be influenced by several factors, including the precision and choice of the brain deformation metric and the accuracy of the concussion diagnosis, which often relies on symptom reporting and lacks a single objective physical measure.
Therefore, recent studies have also adopted a bottom-up approach that links tissue-level deformation metrics measured in controlled \textit{ex vivo} cell culture and \textit{in vivo} animal experiments to cellular injury and cell death \cite{bar2016strain, estrada2021neural, gonzalez2024cortical, hajiaghamemar2020head}, thereby establishing a physics-based link between the deformation metrics, cellular injury, and ultimately to TBI outcomes. However, the full chain linking cellular injury to the probability of concussion remains an active area of investigation. 

To evaluate injury risk based on brain deformation, we focus here on MPS, which was the most consistently reported deformation-based metric in the soccer heading computational studies (Section \ref{strain}).  We consider three different MPS-based injury risk metrics as shown in Figure \ref{InjuryRisk}b.  The first injury risk metric is from Kleiven et al. \cite{kleiven2007predictors}, who correlated NFL (National Football League) concussion data with computed MPS (100th percentile) in different regions of interest (ROI) of the KTH head model. 
Here, we apply the injury risk curve derived for the gray matter. The second MPS-based injury risk curve is from Takhounts et al. \cite{takhounts2013development}, who used animal brain injury data and correlated MPS (100th percentile) from the SIMon and GHBMC human head models to estimate injury risk. 
The final MPS-based injury risk curve is from Hajiaghamemar et al. \cite{hajiaghamemar2020head}, who developed injury risk curves based on spatial correlations between the 95th percentile MPS and histopathological markers of traumatic axonal injury (AIS 4) in porcine models. 

\subsubsection{Injury Risk Analysis based on Brain Deformation}
Figure \ref{InjuryRisk}c shows the probability of brain injury based on Kleiven's injury risk curve \cite{kleiven2007predictors} using the peak MPS computed from the computational soccer studies (Section \ref{strain}) \cite{huber2024finite, perkins2022assessment, barnes2024investigation, filben2021header,brooks2021purposeful}. 
The injury risk is estimated to be below 20$\%$ for all demographics and header types, 
which is consistent with the kinematics-based injury risk estimated from the mouthguard studies (Figure \ref{Combined BrIC}).  
It should be noted that the 95th percentile MPS is reported in most computational studies whereas the injury risk curve from Kleiven et al. \cite{kleiven2007predictors} is based on the 100th percentile MPS; therefore, the actual estimated injury risk from Kleiven et al. may be higher. 
The peak MPS values from the computational studies were below 10\%, which is lower than the values associated with TBI risk from the Takhounts et al. \cite{takhounts2013development} and Hajiaghamemar et al. \cite{hajiaghamemar2020head} studies. Given the variability in the injury risk estimates, these results highlight the need for further computational studies and continued development of brain deformation-based injury risk curves.


\begin{figure}[htbp]
    \centering
    \includegraphics[width=1.\textwidth]{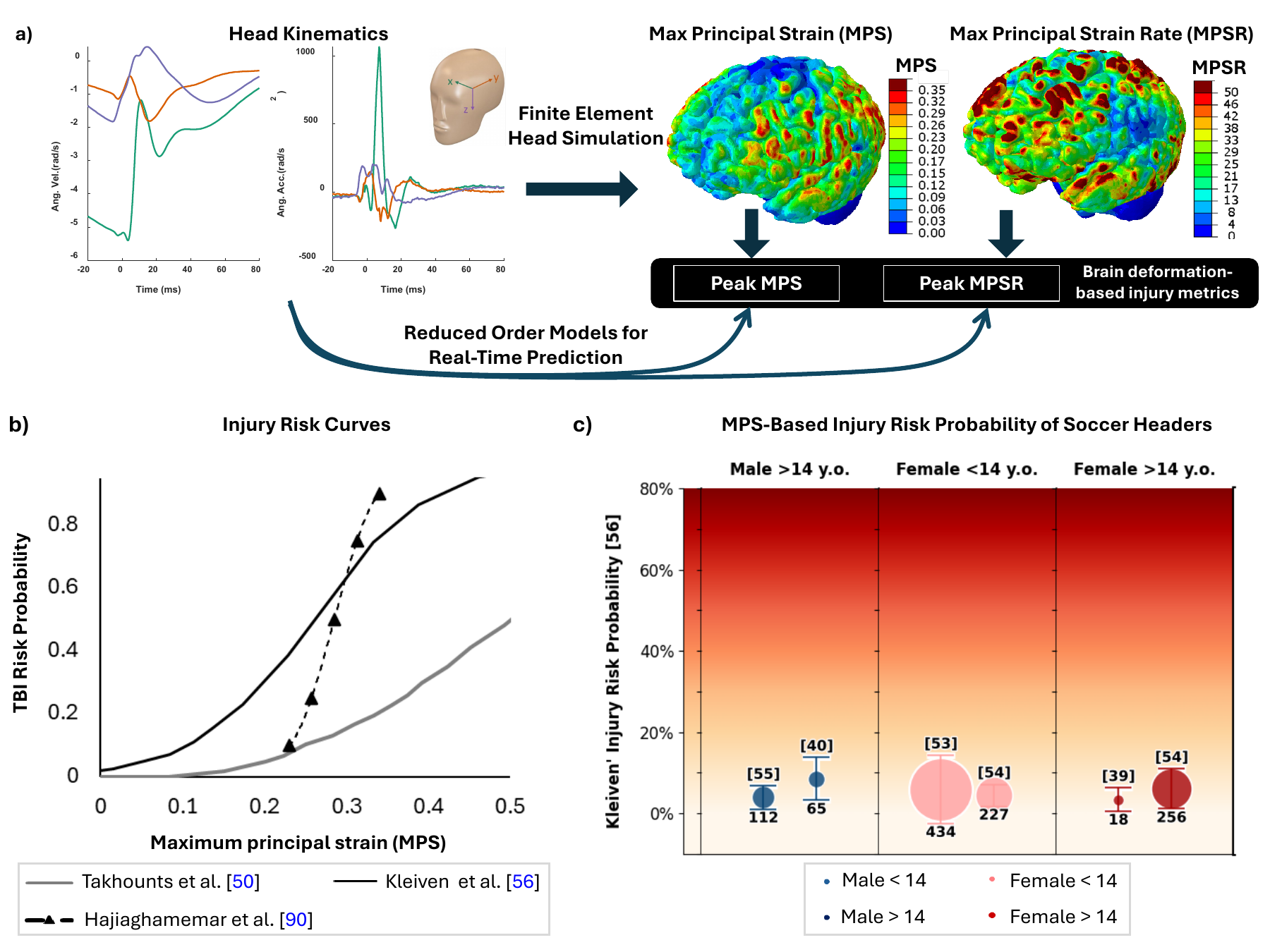}
    \caption{a) Brain tissue deformation measures, such as peak MPS and MPSR, can be estimated from finite element soccer header simulations using the recorded head kinematics as boundary conditions. Real-time estimates of these measures can also be obtained from reduced-order models, idealized continuum models, and machine learning methods. 
    b) Several MPS-based injury risk curves from the literature \cite{takhounts2013development,kleiven2007predictors,hajiaghamemar2020head}.
    c) Estimated brain injury risk associated with soccer heading using the injury risk curve by Kleiven et al. \cite{kleiven2007predictors}.}
    \label{InjuryRisk}
\end{figure}

\section{Limitations of Current Studies}\label{lim}

While the literature on soccer heading and brain biomechanics has expanded in recent years, it remains constrained by several key limitations. These limitations span methodological design, demographic representation, and data reporting. 

\subsection{Sample Size and Demographic Bias}\label{subsec10}
Most available studies draw on small, convenience-based cohorts, which limits statistical power and increases the likelihood of sampling bias. Participants are disproportionately female, adult athletes, leaving youth and male players underrepresented. As a result, findings may not generalize across the broader soccer population.

\subsection{Sensor Technology and Measurement Error}\label{subsec11}
Studies use a wide range of sensor systems: mouthguards, headbands, and skin patches, each with unique accuracy profiles. Headband and skin patch sensors consistently overestimate the true head kinematics of an impact due to their tendency to partially decouple from the skull due to soft tissue movement. 
Studies have shown that the correlation coefficient with the reference sensor data can be as low as 0.58, 0.42, and 0.63 for PLA, PAV, and PAA, respectively, during typical soccer heading for headbands \cite{tripathi2025laboratory,patton2021,hanlon2010validation}. While skin-patches have performed well in laboratory tests (correlation coefficient = 1 for PAV), 
the performance deteriorates on human tests, with the Positive Predictive Value (PPV) for detecting an impact event to be as low as 0.15 \cite{kieffer2020two}. The time history agreement, measured using normalized root mean square error (NRMSE), can be as high as 0.89 and 0.61 in PRV and PRA for headbands \cite{tripathi2025laboratory,patton2021,hanlon2010validation}, and 0.4, 0.76, and 2.3 for PLA, PRV and PRA for skin patch sensors \cite{wu2016vivo}. 
In contrast, several mouthguard sensors have been shown to offer better fidelity to the motion experienced by the head (correlation coefficient $>$ 0.9 for PLA, PAV, and PAA; and PPV $>$ 0.75) \cite{stitt2021laboratory, liu2020validation,kieffer2020two}. Differences in placement, calibration, and filtering methods introduce further variability and hinder cross-study comparisons. 

\subsection{Data reporting and accessibility}
Different studies present data in different formats, e.g., some studies report mean $\pm$ standard deviation, while others report median and inter-quartile range. While missing information from some previous studies was obtained, other studies could not be included in the analyses due to inaccessibility of the data. Furthermore, many  injury metrics require peak angular acceleration, peak angular velocity, impact duration, or the full time histories of each impact.  The lack of these data limit the ability to perform more comprehensive injury analyses.

\subsection{Field vs. Laboratory Conditions}\label{subsec12}
Laboratory reconstructions allow for precise control and verification of impacts but often lack ecological validity, as they cannot fully replicate the dynamics of live match play. Conversely, field studies capture real-world exposures but face challenges in verifying true head contact events and controlling for confounding variables. This trade-off leaves gaps in the ability to translate findings across settings. 

\subsection{Variability in Exposure Definitions}\label{subsec13}
What constitutes a “head impact” varies considerably across studies. Some define exposures strictly as direct ball-to-head contacts, while others include incidental or body-to-body contacts involving head motion. These definitional differences complicate efforts to compare impact rates, magnitudes, and cumulative exposures across cohorts. 

\subsection{Limited Longitudinal and Outcome Data}\label{subsec14}
All the reviewed studies on head kinematics and brain strain reported data without linking exposures to clinical outcomes. Longitudinal studies tracking soccer players across seasons, or into later life, are scarce. Without outcome data, it remains unclear how observed impact frequencies and magnitudes translate into meaningful brain health risks. This gap is especially critical in the context of repetitive sub-concussive impacts. Even if individual headers fall below commonly cited injury thresholds, the cumulative effect of hundreds or thousands of such impacts over a career may still contribute to neurophysiological changes or long-term neurodegenerative risk \cite{montenigro2017cumulative}. Current kinematic datasets cannot resolve these cumulative effects, leaving a major uncertainty in interpreting the clinical significance of soccer heading exposure.

\subsection{Conflation of Concussion and mTBI in Injury Risk Metrics}\label{subsec15}Another limitation of this meta-analysis 
is the use of concussion-based injury risk criteria as a proxy for mTBI. Although these terms are used nearly interchangeably, they differ by definition. A concussion is a type of traumatic brain injury, clinically diagnosed by a set of possible signs and symptoms \cite{mccrory2017consensus}. On the other hand, mTBI is characterized by neurobiological alterations that are mechanically induced, 
which may occur without overt symptoms \cite{silverberg2023american}. A majority of the current injury risk functions, including some of those selected for this review, are based on clinically diagnosed concussion datasets \cite{rowson2013brain,takhounts2013development}. Because of this, the functions reflect only the subset of brain injuries that were symptomatic and recognized, rather than the full spectrum of underlying tissue-level damage. 

This distinction becomes important to recognize when analyzing the broad spectrum of heading intensities found in the sport of soccer. While most impacts are classified as sub-concussive \cite{kroshus2015concussion,o2014evaluation}, some of these may still induce measurable brain deformation (MPS, CSDM) and potentially contribute to cumulative neurological changes over time. Therefore, the current injury risk functions systematically underestimate the occurrence of brain injury by excluding subclinical asymptomatic cases \cite{fawzi2026toward}.

\section{Future Directions}\label{future}
Future research on soccer heading and brain biomechanics should prioritize methodological rigor and broader demographic inclusion. Standardization of sensor technologies and analytic pipelines is critical for comparability across studies, particularly in quantifying head impact kinematics. Advances in wearable sensor accuracy and video-verified event classification will help reduce measurement error and clarify the true exposure profile of players. In this context, alignment with the recently published Consensus Head Acceleration Measurement Practices (CHAMP) guidelines will be essential. CHAMP outlines the best practices for validating wearable devices, reporting measurement protocols, and processing head-acceleration data, emphasizing independent laboratory validation using ATDs, standardized reference sensors, and transparent data-handling procedures \cite{gabler2022consensus}. Adopting these practices will substantially improve the reproducibility and reliability of future wearable-sensor studies in soccer.

Equally important is the standardization of injury risk metrics. While raw kinematic measures such as PLA, PAV, and PAA describe the magnitude of impacts, their clinical interpretation depends on how they are translated into risk through criteria like BrIC or probability functions derived from impact databases. At present, studies apply different thresholds, models, and scaling factors, making cross-study comparisons difficult and sometimes misleading. Establishing consensus on which injury criteria to use, how to report them, and how to validate them against clinical outcomes are essential for future comparability and clinical relevance. It is also important that the full kinematic time history data of each impact is provided to enable the calculation of these injury metrics.

Demographic gaps must also be addressed. Most existing work has focused on adult female cohorts, leaving male and youth athletes underrepresented. Future studies should deliberately recruit across sex and age groups to better characterize developmental and sex-based differences in head impact exposure and outcomes. Longitudinal designs following players across multiple competitive seasons would further illuminate cumulative effects that are currently poorly understood. 

Finally, future research should move beyond laboratory simulations and small training-based cohorts toward large-scale, live match-play studies. Integrating kinematic data with emerging neuroimaging, blood biomarkers, and neurocognitive testing will be essential for linking exposure metrics such as PLA, PAV, and PAA 
to clinical relevance. Such interdisciplinary approaches will help bridge the gap between biomechanical measurement and the long-term neurological safety of soccer players. 


\backmatter


\section*{Declarations}

\begin{itemize}
\item \textbf{Funding:} The authors gratefully acknowledge the support from the University of Wisconsin-Madison Office of the Vice Chancellor for Research (OVCR) and the Athletic Department. Funding for this award has been provided through Big 10 Athletic Media Revenue (136-AAI3375). The authors also acknowledge funding from the U.S. Office of Naval Research under award N00014-24-1-2415 through Dr. Timothy Bentley.

\item \textbf{Competing interests:} The authors have no competing interests to declare that are relevant to the content of this article. 

\item \textbf{Author Contributions:} All authors contributed to the conception and design of the review. The literature search was performed by Christopher Lewis, Anu Tripathi, and Rika Carlsen; the data were synthesized by Christopher Lewis; and statistical analyses were performed by Anu Tripathi. The original draft was prepared by Christopher Lewis, and revised by Anu Tripathi and Rika Carlsen. All authors reviewed the manuscript and approved the final manuscript.

\item \textbf{Data Availability:} All analyzed data will be made available upon request.
\end{itemize}

\nocite{*}
\bibliography{sn-bibliography}

\end{document}